# Pattern selection as a nonlinear eigenvalue problem

P. Büchel and M. Lücke

Institut für Theoretische Physik,
Universität des Saarlandes, D-66041 Saarbrücken, Germany

**Abstract.** A unique pattern selection in the absolutely unstable regime of driven, nonlinear, open-flow systems is reviewed. It has recently been found in numerical simulations of propagating vortex structures occuring in Taylor-Couette and Rayleigh-Bénard systems subject to an externally imposed through-flow. Unlike the stationary patterns in systems without through-flow the spatiotemporal structures of propagating vortices are independent of parameter history, initial conditions, and system length. They do, however, depend on the boundary conditions in addition to the driving rate and the through-flow rate. Our analysis of the Ginzburg-Landau amplitude equation elucidates how the pattern selection can be described by a nonlinear eigenvalue problem with the frequency being the eigenvalue. Approaching the border between absolute and convective instability the eigenvalue problem becomes effectively linear and the selection mechanism approaches that of linear front propagation.

## 1 Introduction

In many nonlinear continuous systems dissipative structures branch off a homogeneous basic state when the external stress exceeds a critical threshold. Examples for these transitions are Taylor-Couette flow, Rayleigh-Bénard convection, binary-fluid convection, flame-front propagation, and some chemical or biological processes [1]. Often, for a fixed configuration of parameters and boundary conditions (BCs) a continuous or discrete family of patterns with different wave numbers is stable. Their stability regime, e. g., a band of wave numbers is usually limited by secondary instabilities. The stable structures within such a band can be generated by appropriately engineered time histories of the control parameters and/or by properly changing the BCs, e. g., the system size. The most intensively investigated examples in this respect are the structures of Taylor vortices [1, 2, 3, 4, 5, 6, 7] in an annulus between concentric cylinders of which the inner one rotates, convective roll vortices in horizontal layers of one-component fluids [1, 8, 9, 10, 11, 5], and binary mixtures heated from below [12, 13].

This multiplicity of solutions of the underlying nonlinear partial differential equations that stably coexist for a fixed configuration of parameters and BCs seems to disappear in open-flow systems: Recent numerical simulations of Rayleigh-Bénard convective roll vortices traveling downstream in an imposed horizontal Poiseuille flow [14, 15, 16, 17] and of Taylor vortices propagating downstream in an axial through-flow [18] showed that the spatiotemporal structure of these propagating vortex (PV) flows is uniquely selected, i.e., independent of parameter history, intial conditions, and system size in the absolutely unstable regime. Therein the pattern starting, e. g., from a spatially localized perturbation can grow in upstream as well as in downstream direction [19]. The structure expansion proceeds until the upstream (downstream) moving front encounters in a finite system the inlet (outlet) and adjusts to the inlet (outlet) BC. The final pattern resulting in such a situation shows a characteristic streamwise profile of the amplitude growing with increasing distance from the inlet and of the wave number, and a characteristic global, i.e., spatially constant oscillation frequency associated with the downstream motion of the pattern.

By contrast, in the convectively unstable regime [20] initial perturbations are blown out of the system — both, the upstream as well as the downstream facing front of the growing structure move downstream. Therefore it requires a persistent perturbation source like, e. g., noise to sustain a vortex pattern in the convectively unstable regime [21]. It should be emphasized that the structural dynamics of pattern formation in the convectively unstable regime at larger through-flow rates and/or smaller driving rates substantially differs from the one reviewed here in the absolutely unstable regime. The latter regime is governed by nonlinear contributions in the balance equations; the resulting patterns are uniquely selected and insensitive to initial conditions, parameter history, and small perturbations. On the other hand, in the convectively unstable regime the growing patterns are sensitive to initial conditions and perturbations.

In this work we review how such a uniquely selected spatiotemporal pattern structure in the absolutely unstable regime can be understood as a nonlinear eigenvalue problem with the oscillation frequency being the eigenvalue and the profiles of pattern intensity and wave number determining the corresponding eigenfunction. Although the PV structures selected by the full hydrodynamic field equations called Navier-Stokes equations (NSE) for short and the Ginzburg-Landau amplitude equation (GLE) approximation differ in a characteristic way the selection mechanism looks similar. The frequency selection seems to result from requiring the spatial variation of pattern envelope and phase to be as small as possible under the imposed inlet/outlet BC which is analogous to ground state properties of the linear stationary Schrödinger equation.

We also show how the pattern selection process is related to the one occurring behind a front or "domain wall" that spatially separates an unstable, homogeneous state from a stable, structured state. Approaching the border between the absolutely and convectively unstable regimes the pattern selection mechanism becomes linear: For driving and through-flow rates on this border line the selected frequency is the one resulting from a linear front whose spatiotemporal behavior is governed by the fastest growing linear mode. The latter is identified by a particular saddle of the complex linear dispersion relation over the complex wave number plane [1]. Since the linear dispersion relations of NSE and GLE deviate from each other for supercritical control parameters, the PV structures selected by NSE or GLE differ in a characteristic way.

## 2   The Systems

We have investigated on the one hand time-dependent, rotationally symmetric vortex structures in the Taylor-Couette system with an externally enforced axial flow [18] and on the other hand straight convection roll vortices in the Rayleigh-Bénard system subject to an externally applied plane horizontal flow with roll axes perpendicular to the through-flow [14, 15, 16, 17]. In both cases a translationally invariant flow state is stable below a threshold of the driving, $T_c(Re)$ [22, 23, 24, 25] of the Taylor number $T$ or $Ra_c(Re)$ [26, 15, 16] of the Rayleigh number $Ra$, respectively, which depend on the through-flow Reynolds number $Re$.

At these critical thresholds the basic state becomes unstable to PV perturbations via an oscillatory instability and in an infinite system a nonlinear PV solution branches off the basic state. We use the relative control parameter

$$\mu = \frac{T}{T_c(Re)} - 1 \quad \text{or} \quad \mu = \frac{Ra}{Ra_c(Re)} - 1 \tag{1}$$

corresponding to

$$\epsilon = \frac{T}{T_c(Re=0)} - 1 \quad \text{or} \quad \epsilon = \frac{Ra}{Ra_c(Re=0)} - 1 \tag{2}$$

to measure the distance from the onset of PV flow for $Re \neq 0$ and of stationary Taylor or convection roll vortex flow for $Re = 0$, respectively. In this notation $\mu_c = 0$, i.e.,

$$\epsilon_c(Re) = \frac{T_c(Re)}{T_c(Re=0)} - 1 \quad \text{or} \quad \epsilon_c(Re) = \frac{Ra_c(Re)}{Ra_c(Re=0)} - 1 \tag{3}$$

is the critical threshold for onset of PV flow and $\mu = \epsilon/(1+\epsilon_c(Re))$. The shear forces associated with the through-flow slightly stabilize the homogeneous basic state, so $\epsilon_c(Re)$ slightly increases with $Re$ [22, 25, 27, 28, 29, 15].

In the remainder of this paper we use the Taylor-Couette system as a representative example when displaying specific equations. But one should keep in mind that our results equally well apply to the problem of propagating convection roll vortices in the Rayleigh-Bénard setup and also to other similar pattern forming systems.

### 2.1 Ginzburg-Landau amplitude equation

To characterize the secondary PV structure we consider the *deviation* of the velocity field from the basic state as the order-parameter field. Close to the bifurcation threshold of PV flow, i.e., for small $\mu$, the flow has the form of a harmonic wave. For example, the axial component of the vortex field in the Taylor-Couette system reads

$$w(r,z;t) = A(z,t)\, e^{i(k_c z - \omega_c t)} \hat{w}(r) + c.c. \tag{4}$$

with a complex amplitude $A(z,t)$ that is slowly varying in the axial direction $z$ and time $t$. Here and in the following all quantities are properly reduced [18, 14] to be non dimensional. The critical wave number $k_c$, frequency $\omega_c$ [23, 24, 22, 25, 27, 28], and eigenfunction $\hat{w}(r)$ [23, 30] depending on the radial coordinate $r$ have been obtained from a linear stability analysis of the basic flow state as functions of $Re$. The complex vortex amplitude $A(z,t)$ is given by the solution of the 1D complex GLE,

$$\tau_0 \left( \dot{A} + v_g A' \right) = \mu(1 + ic_0) A + \xi_0^2 (1 + ic_1) A'' - \gamma(1 + ic_2) |A|^2 A. \tag{5}$$

Dot and primes denote temporal and spatial derivatives in the $z$-coordinate, respectively. As a consequence of the system's invariance under the combined symmetry operation $\{z \to -z, Re \to -Re\}$ the coefficients $\tau_0, \xi_0^2, \gamma$ are even in $Re$ while the group velocity $v_g$ and the imaginary parts $c_0, c_1, c_2$ are odd in $Re$ [16, 22]. An analogous expression of (4) and GLE also holds for convective roll vortices in the Rayleigh-Bénard system.

It should be noted that the control parameter range of $\mu$ over which (4) gives an accurate description, say, on a percent level is indeed very small in the Taylor-Couette system: The asymmetry between radial in- and outflow intensities rapidly grows with $\mu$ and causes higher axial Fourier contributions $\sim e^{inkz}$ [31, 32, 33] to the velocity field that are discarded in the $\mu \to 0$ asymptotics of the GLE approximation (4). In the Rayleigh-Bénard system the situation is more favorable for a GLE approximation. In any case however, the modulus of the first Fourier mode of the vortex structures agrees in both systems for $Re = 0$ as well as for $Re \neq 0$ quite well with the one predicted by the GLE.

### 2.2 Absolute and convective instability

For small $\epsilon$ and $Re$ the control parameter plane is divided into three stability regimes - cf. Fig. 1 - characterized by different growth behavior of *linear* perturbations of the basic flow state. In the presence of through-flow one has to distinguish [19] between the spatiotemporal growth behavior of spatially localized perturbations and of spatially extended ones. Below the critical threshold $\epsilon_c(Re)$ for onset of PV flow (dashed line in Fig. 1) any perturbation, spatially localized as well as extended, decays. This is the parameter regime of absolute stability of the basic state.

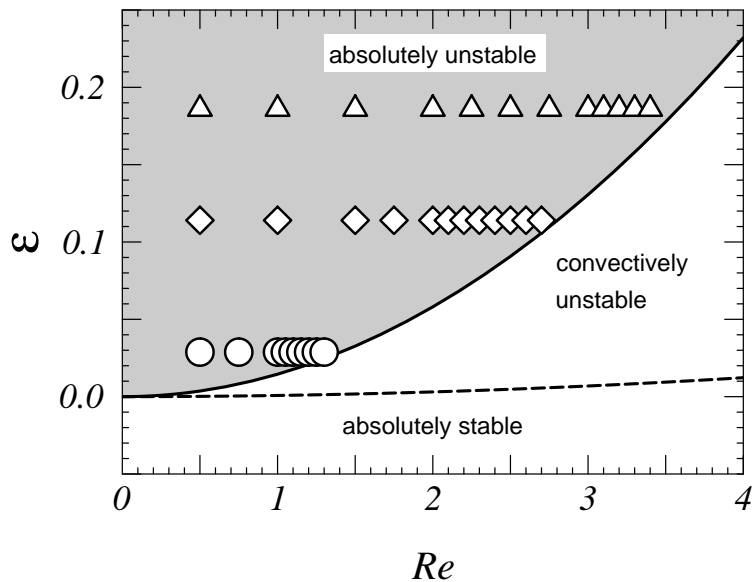

**Fig. 1.** Stability domains of the basic state. Numerical simulations of the Taylor-Couette system have been performed for the parameters marked by symbols. Dashed line is the critical threshold for onset of extended PV flow and full line the boundary (6) between absolute and convective instability.

For $\epsilon > \epsilon_c$ extended perturbations can grow. A spatially localized perturbation, i.e., a wave packet of plane wave perturbations is advected in the so-called convectively unstable parameter regime faster downstream than it grows — while growing in the comoving frame it moves out of the system [19, 20, 21, 34]. Thus, the downstream as well as the upstream facing intensity front of the vortex packet move into the same direction, namely, downstream. A spatially localized perturbation is blown out of any system of finite length and the basic state is reestablished. It therefore requires a persistent perturbation source like, e. g., noise to sustain a vortex pattern in the convectively unstable regime [21].

In the absolutely unstable regime (shaded region in Fig. 1) a localized perturbation grows not only in the downstream direction but it grows and spatially expands also in the upstream direction until the upstream propagating front encounters the inlet in a finite system. The final pattern resulting in such a situation shows in the downstream direction a characteristic intensity profile under which the PV flow develops with increasing distance from the inlet. Approaching the boundary between absolute and convective instability the growth length of the PV structure diverges in the absence of any perturbation source and the PV pattern is blown out of the system. However, in the presence of noise there is a transition to a noise sustained structure with a characteristic finite growth length depending on noise properties and control parameters.

Within the framework of the amplitude equation the boundary (full line in Fig. 1) between absolute and convective instability is given by [21]

$$\mu^c_{conv} = \frac{\tau_0^2 v_g^2}{4\xi_0^2 (1 + c_1^2)} \qquad (6)$$

corresponding to $\epsilon^c_{conv} = \epsilon_c + (1 + \epsilon_c)\mu^c_{conv}$. Thus the absolutely unstable regime (shaded region in Fig. 1) is characterized by $\mu > \mu^c_{conv}$ or, equivalently, by the reduced group velocity

$$V_g = \frac{\tau_0}{\xi_0 \sqrt{(1 + c_1^2) \mu}} v_g = 2\sqrt{\frac{\mu^c_{conv}}{\mu}} \qquad (7)$$

being smaller than 2. It should be mentioned that the GLE approximation (6) of $\mu^c_{conv}$ describes the boundary between absolute and convective instability resulting from the NSE [28, 15, 16, 17, 35] very well for the small Reynolds numbers considered here.

When comparing results obtained for different $\mu$ and $Re$ from numerical simulations of the NSE with those following from the GLE we found it sometimes advantageous to present them as functions of the reduced group velocity $V_g$ (7). The deviation

$$2 - V_g = 2 - 2\sqrt{\mu^c_{conv}/\mu} \qquad (8)$$

is a scaled distance from the boundary between the absolutely and convectively unstable regimes.

## 3   Propagating Vortex Patterns

We have performed numerical simulations of the full 2D NSE for rotationally symmetric Taylor vortices and straight convective roll vortices in systems of finite length [18, 14, 15, 16, 17]. They revealed in the absolute unstable regime for various end conditions patterns of vortices propagating downstream under a stationary intensity envelope after transients have died out. The oscillation frequency $\omega$ of the PV flows is spatially constant while the local wave number $k$, the phase velocity $v_p = \omega/k$, and the vortex flow intensity varies. The frequency and the spatial variation of the PV pattern depends on the control parameters and on the BCs but not on parameter history or initial conditions.

For simplicity we here only discuss vortex suppressing BCs that are realized by imposing the homogeneous basic state at the inlet ($z = 0$) and outlet ($z = \Gamma = 50$) of the system. In the GLE approximation this amounts to requiring $A = 0$ at the inlet and outlet. To illustrate the global properties of the flow patterns we present in Fig. 2 a hidden-line plot of the axial velocity field $w$ in the Taylor-Couette system. Thin lines show snapshots of $w$ at the radial position $r = r_1 + 0.225$ obtained from the NSE at successive, equidistantly spaced times. After transients have died out the vortices propagate downstream under a stationary intensity envelope (thick line) determined by the temporal extrema of $w(r_1 + 0.225; t)$ at any z-position. The intensity variation of PV flow under the fronts causes there a spatial variation in the local wavelength $\lambda(z)$ and in the phase velocity $v_p(z) = \omega \lambda(z)/2\pi$ with the oscillation frequency $\omega$ of PV flow being globally constant [18, 14].

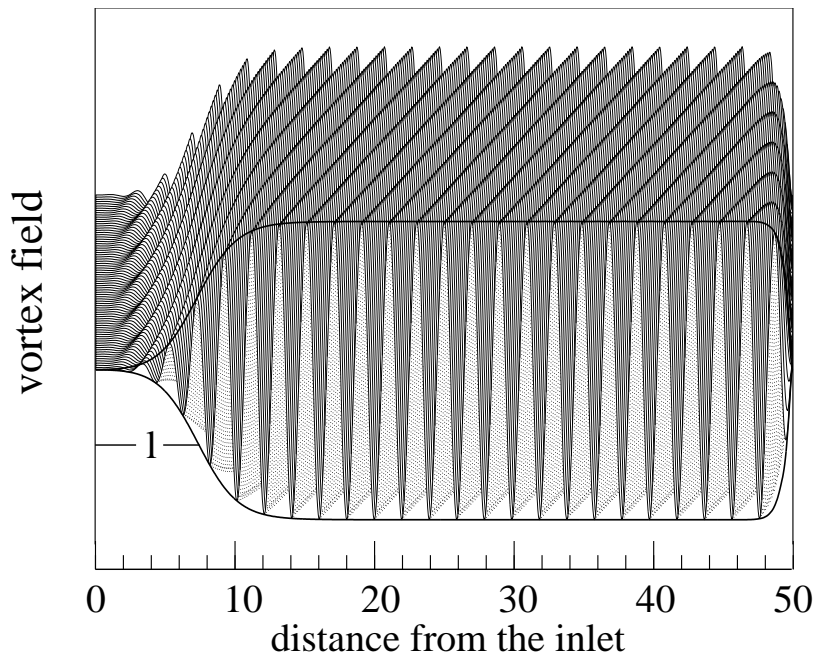

**Fig. 2.** Spatiotemporal structure of PV flow. Thin lines show vertically displaced snapshots of the vortex velocity field at successive, equidistantly spaced times. Thick lines show the stationary envelope. The BC at the inlet (left) and outlet (right) suppresses any vortex flow there.

For further characterization of the PV flow structure we found a *temporal* Fourier decomposition of the time-periodic fields, e.g.,

$$w(r, z; t) = \sum_n w_n(r, z) e^{-in\omega t} \qquad (9)$$

to be useful. Note that the long-time solution of the GLE

$$A(z, t) = R(z) e^{i[\varphi(z) - \Omega t]} \qquad (10)$$

8      P. Büchel and M. Lücke

oscillates harmonically with frequency $\Omega$ under a stationary envelope

$$R(z) = |A(z,t)| \tag{11}$$

after transients have died out. Therefore, the GLE fields contains no temporal Fourier mode other than $n = 1$ whereas the solution of the full NSE fields have higher harmonics and a stationary contribution of a zeroth fourier mode [18, 14].

Sufficiently away from the inlet and outlet we observe a bulk region of nonlinear saturated PV flow with spatially uniform amplitude and wavelength. In the bulk region the *temporal* modes obtained for finite through-flow from the NSE grow for small $\mu$ proportional to $\mu^{n/2}$ with relative corrections proportional to $\mu$. Thus, they show the same growth behavior with $\mu$ that the *spatial* Fourier modes [31, 36, 37, 32, 33] of stationary vortices without through-flow show as a function of $\epsilon$. The reason is that all fields in the PV state have the form of propagating waves

$$f_b(r, z; t) = f_b(r, z - \frac{\omega}{k_b}t) \tag{12}$$

in the bulk region - denoted by a subscript $b$.

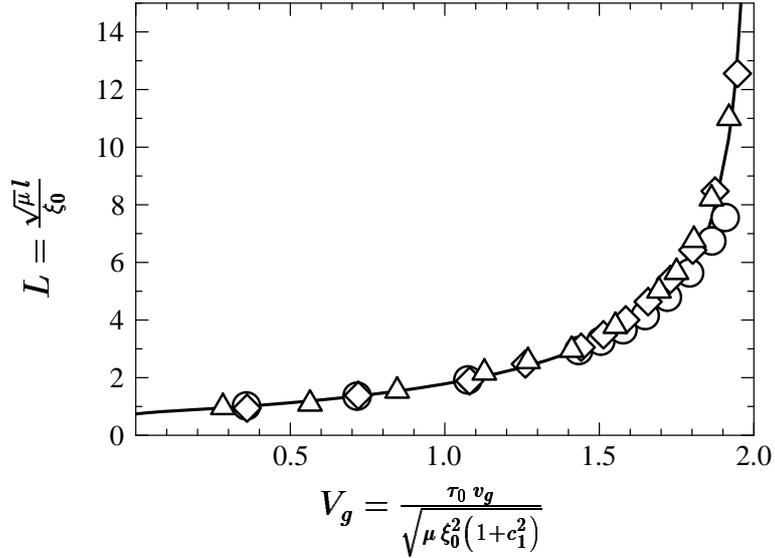

**Fig. 3.** Scaled growth length $L$ (13) of PV structures versus scaled group velocity $V_g$ (7). Symbols represent lengths obtained from the NSE for different combinations (cf. Fig. 1) of $Re$ and $\epsilon = 0.0288$ (circles), $\epsilon = 0.114$ (diamonds), and $\epsilon = 0.186$ (triangles). The line shows the scaling behavior of the GLE.

With increasing through-flow this bulk pattern is pushed further and further downstream. The growth length $l$ from the inlet over which the amplitude reaches half its saturation value depends on $\mu$ and $Re$. It increases and finally diverges when the control parameters $\mu$, $Re$ approach the absolute-convective instability border $\mu_c^{conv}$, that is when $V_g$ (7) reaches 2. In Fig. 3 we compare the scaled growth length

$$L = \sqrt{\mu}\, l/\xi_0 \qquad (13)$$

computed from the amplitude equation (solid line) with results from the NSE (open symbols) for various combinations of $\epsilon$ and $Re$. Due to the scaling property of the GLE [15, 16, 17] keeping in mind the smallness of the imaginary parts $c_i$ all values for $L$ obtained by the GLE fall onto one curve for different control parameters. The NSE results lie very close to the GLE curve. This also holds for propagating convective roll vortices [14, 15, 16, 17].

## 4 Pattern selection

Our investigations of the PV structures in the Taylor-Couette system [18] and in the Rayleigh-Bénard system [14, 15, 16, 17] have shown that the through-flow causes a unique pattern selection such that the selected structure is independent of history and initial conditions and depends only on the control parameters and BCs. Thus, the flow causes possible wave numbers within the Eckhaus-stable band of stationary Taylor or convection vortex patterns without through-flow to collapse to only one uniquely selected PV structure for nonvanishing through-flow rates.

### 4.1 Pattern selection within the GLE

Here we elucidate how the pattern selection mechanism of the GLE can be understood as a nonlinear eigenvalue/boundary-value problem where the frequency of the PV pattern is the eigenvalue. Thus, the selected PV structure is characterized and determined by the combination of eigenvalue and corresponding eigenfunction.

**The eigenvalue problem** We look for solutions of the GLE of the form (10);

$$A(z,t) = a(z) e^{-i\Omega t} = R(z) e^{i[\varphi(z) - \Omega t]}, \qquad (14)$$

with stationary envelope $R(z)$, stationary wave number

$$q(z) = k(z) - k_c = \varphi'(z), \qquad (15)$$

and constant frequency

$$\Omega = \omega - \omega_c. \qquad (16)$$

We have numerically solved the GLE and verified that its long-time solution is indeed of the form (10). Inserting this solution ansatz into the GLE one obtains the nonlinear eigenvalue problem

$$\tau_0 \left(-i\Omega a + v_g a'\right) = \mu \left(1 + ic_0\right) a + \xi_0^2 \left(1 + ic_1\right) a'' - \gamma \left(1 + ic_2\right) |a|^2 a \quad (17)$$

or, equivalently,

$$i\tau_0(-\Omega + v_g q)R + \tau_0 v_g R' = \mu \left(1 + ic_0\right)R + \\ \xi_0^2(1 + ic_1)(R'' - q^2 R + iq'R + 2iqR') - \gamma(1 + ic_2)R^3 \quad (18)$$

as a solvability condition with $\Omega$ and $a(z)$ being the eigenvalue and associated eigenfunction, respectively. Here we consider the BCs $R_{in,out} = 0$ at the inlet and outlet. These requirements fix four BCs — namely, $\Re A_{in,out} = \Im A_{in,out} = 0$ — which are necessary to solve the GLE.

We are interested in solutions for which the variation of $R(z)$ and in particular of $q(z)$ is "as small as possible". Such a solution type seems to be connected with the eigenvalue $\Omega$ that is closest to zero. While we cannot make statements about the eigenvalue spectrum our numerical solutions of the GLE indicate that initial conditions always evolve into a pattern with smooth amplitude $a(z)$ and small frequency $\Omega$. In that sense the eigenvalue problem (17) resembles ground state problems of the Schrödinger equation — increasing the spatial variation of the wave function (amplitude $a$) implies higher kinetic energy (frequency $\Omega$).

**Inlet behavior** While the eigenfrequency $\Omega$ is in general determined by *global* properties and not by the *local* variation of the eigenfunctions $R(z)$ and $q(z)$, say, at the inlet or in the bulk, it is informative and useful for our further discussion to list below relations between $\Omega$ and structural properties at the inlet and in the bulk of the system. From (18) one finds that $R = 0$ at the inlet and outlet implies the relation

$$R''_{in,out} = [\frac{\tau_0 v_g}{\xi_0^2 (1 + ic_1)} - 2iq_{in,out}]R'_{in,out} \quad (19)$$

at $z = 0, \Gamma$. Hence, whenever the modulus grows with finite slope, $R'_{in,out} \neq 0$, the BC $R = 0$ fixes the wave number at the inlet *and* outlet to the value $q_{in} = q_{out}$,

$$\xi_0 q_{in,out} = -\frac{1}{2} c_1 \frac{\tau_0 v_g}{\xi_0 (1 + c_1^2)} = -\frac{c_1}{\sqrt{1 + c_1^2}} \sqrt{\mu^c_{conv}} \quad (20)$$

which depends only on $Re$. This value ensures that the imaginary part of the expression in the square brackets in (19) is zero. For $Re = 0$ one obtains $q_{in,out} = 0$ as Cross *et al.* [38]. In addition the condition $R = 0$ yields via the real part of (19) two relations

$$R''_{in,out} = \frac{\tau_0 v_g}{\xi_0^2 (1 + c_1^2)} R'_{in,out} \quad (21)$$

between the different slopes $R'_{in} \neq R'_{out}$ and curvatures $R''_{in} \neq R''_{out}$ of the modulus at the two respective boundaries that also hold independently of $\mu$.

Furthermore, one can derive a relation between the eigenfrequency $\Omega$, the wave number $q_{in,out}$, and its slope $q'_{in,out}$ at the inlet and outlet. To that end we consider the spatial derivative of (18) at the inlet and outlet with $R = 0$:

$$i\tau_0(\Omega - v_g q) + \mu(1 + ic_0) - \xi_0^2(1 + ic_1)q^2 - [\tau_0 v_g - 2iq\xi_0^2(1 + ic_1)]\frac{R''}{R'}$$
$$= \xi_0^2(1 + ic_1)(\frac{R'''}{R''} + 3iq'). \quad (22)$$

This allows one to solve for $\frac{R'''}{R''}$ and $q'$ separately. Using (20) and (21) one obtains the relation

$$\tau_0 \Omega = (c_1 - c_0)\mu + \tau_0 v_g q_{in,out} - 3(1 + c_1^2)\xi_0^2 q'_{in,out}. \quad (23)$$

Note that this equation implies $q'_{in} = q'_{out}$.

The relation (23) also demonstrates the nonlocality of the pattern selection mechanism and of the eigenvalue problem. At first sight one might be tempted to infer from the above relation between $\Omega$ and $q_{in}, q'_{in}$ that the latter two fix the frequency. However, Eq. (23) holds equally well for $q_{out}, q'_{out}$ and the outlet properties do not fix $\Omega$ either. So the correct interpretation of Eq. (23) is that the eigenvalue $\Omega$, being the characteristic global signature of the pattern, fixes the local quantities $q'_{in,out}$ for the given BC.

**Bulk behavior** The wave number $q(z)$ away from the inlet and outlet in general differs from $q_{in}(= q_{out})$. In order to obtain the full spatial profiles of the eigenfunctions $R(z)$ and $q(z)$ belonging to the eigenvalue $\Omega$ one has to solve the eigenvalue problem (17) with the BC $R_{in,out} = 0$. Alternatively one can solve numerically the time-dependent GLE.

When the system size and the control parameters are such that the PV pattern forms a bulk part with a homogeneous modulus $R_b$ and wave number $q_b$, i. e., where

$$R'_b = R''_b = q'_b = 0, \quad (24)$$

then the dispersion relation

$$\tau_0 \Omega = (c_2 - c_0)\mu + \tau_0 v_g q_b + (c_1 - c_2)\xi_0^2 q_b^2 \quad (25)$$

provides a relation between the frequency eigenvalue $\Omega$, the bulk wave number $q_b$, and the bulk modulus

$$R_b^2 = \frac{\mu - \xi_0^2 q_b^2}{\gamma}. \quad (26)$$

The equation (25) establishes together with (23) also a relation between $q_b$ on the one hand and $q_{in} = q_{out}$ and $q'_{in} = q'_{out}$ on the other hand.

Approaching the convective instability boundary (6) we found that $q'_{in}$ decreases to zero so that the frequency eigenvalue becomes

$$\tau_0 \Omega^c_{conv} = -(c_0 + c_1)\mu^c_{conv}. \tag{27}$$

Then the bulk wave number approaches according to (27) and (25) the limiting value

$$\xi_0(q_b)^c_{conv} = -\frac{\sqrt{1+c_1^2} \overset{(+)}{-} \sqrt{1+c_2^2}}{c_1 - c_2}\sqrt{\mu^c_{conv}}, \tag{28}$$

while the inlet wave number $q_{in}$ is given by (20). Due to the very small imaginary parts $c_i$ the solution (28) with the plus sign has to be discarded, since it corresponds to unphysically large wave numbers.

**Comparison with front propagation** It is suggestive to compare properties of PV patterns like the one in the Fig. 2 with those behind an upstream facing front that connects in an infinite system the basic state at $z = -\infty$ with the developed PV state at $z = +\infty$. A *linear* growth analysis of the GLE along the lines of [Sec. VI B 3;1] shows that the far tail of a *linear* front of PV perturbations moves with the front velocity

$$v_F = v_g - 2\frac{\xi_0}{\tau_0}\sqrt{\mu(1+c_1^2)}. \tag{29}$$

In the convectively unstable regime $0 < \mu < \mu^c_{conv}$ the front propagates downstream so that $v_F > 0$. In the absolutely unstable regime $\mu > \mu^c_{conv}$ it moves upstream, i.e., $v_F < 0$. And at the boundary $\mu = \mu^c_{conv}$ the front is stationary — $v_F = 0$. We compare the properties of such a stationary front at $\mu = \mu^c_{conv}$ with those of the solution $A(z,t)$ (10) with stationary modulus in a semi-infinite system ($0 \leq z < \infty$). The reason for restricting the comparison to $\mu^c_{conv}$ is that an upstream moving front in the absolutely unstable regime will be pushed against the inlet whence the spatiotemporal structure of the free front gets modified by the inlet.

In the linear part of the stationary front the flow amplitude

$$A(z,t) \sim e^{i[Q_s z - \zeta(Q_s)t]} \tag{30}$$

varies with a wave number $q_F = \Re Q_s$, a spatial decay rate $\Im Q_s$, a temporal growth rate $\Im \zeta(Q_s) = 0$, and an oscillation frequency $\Omega_F = \Re \zeta(Q_s)$. Here

$$\xi_0 Q_s = -\frac{1}{2}\frac{i}{1+ic_1}\frac{\tau_0}{\xi_0}v_g = -\frac{i+c_1}{\sqrt{1+c_1^2}}\sqrt{\mu^c_{conv}} \tag{31}$$

is the saddle position of the complex dispersion relation

$$\tau_0 \zeta(Q,\mu) = \tau_0 v_g Q + i(1+ic_0)\mu - i(1+ic_1)\xi_0^2 Q^2 \tag{32}$$

of the linear GLE in the complex $Q$-plane determined by the condition

$$\left.\frac{d\zeta(Q)}{dQ}\right|_{Q_s} = 0 \qquad (33)$$

for $v_F = 0$, i.e., at $\mu = \mu_{conv}^c$.

Thus, the stationary front resulting from the linear GLE is characterized by the local wave number in the far tail

$$\xi_0 q_F(v_F = 0) = -\frac{c_1}{\sqrt{1+c_1^2}}\sqrt{\mu_{conv}^c} \qquad (34)$$

and the oscillation frequency

$$\tau_0 \Omega_F(v_F = 0) = -(c_0 + c_1)\mu_{conv}^c. \qquad (35)$$

So the inlet wave number $q_{in}$ (20) agrees with $q_F$ (34) and $\Omega_{conv}^c$ (27) coincides with $\Omega_F$ (35). Thus, right at the border $\mu_{conv}^c$ of the absolutely unstable regime the frequency eigenvalue of the PV pattern growing in the downstream direction from the inlet at $z = 0$ agrees with the oscillation frequency selected by the stationary, *linear* front. But the PV front pattern develops from the basic state with an exponential growth rate $\kappa$ of the modulus. This leads for $R \to 0$ to $R' = \kappa R, R'' = \kappa^2 R, \ldots \to 0$ for $z \to -\infty$. On the other hand, the PV flow intensity in the semi-infinite system in general drops to zero at the inlet with finite $R'$ and $R''$. Only at the border line $\mu = \mu_{conv}^c$ the difference disappears since there $R', R'', \ldots \to 0$ for $z = 0$. In the absolutely unstable regime, $V_g < 2$, the frequency eigenvalues $\Omega = \omega - \omega_c$ of the GLE lie slightly above the limiting value $\Omega_{conv}^c$ at the border of the absolutely unstable regime. The half-tone symbols in Fig. 4 show for three different $\epsilon$ how $\Omega$ obtained numerically by integrating the GLE approaches with increasing $Re$ the front frequency $\Re\zeta(Q_s) = \Omega_F(v_F = 0) = \Omega_{conv}^c$ (dashed line). For convenience, the merging points with this limit line are marked for the three $\epsilon$-values investigated here by small half-tone symbols.

Inserting at $\mu_{conv}^c$ the linear front frequency $\Omega_F(v_F = 0) = \Omega_{conv}^c$ into the bulk dispersion relation (25) leads to an expression for the bulk wave number far behind the front. For $v_F = 0$ one obtains

$$\xi_0(q_b)_F = \xi_0(q_b)_{conv}^c = -\frac{\sqrt{1+c_1^2} \overset{(+)}{-} \sqrt{1+c_2^2}}{c_1 - c_2}\sqrt{\mu_{conv}^c}. \qquad (36)$$

As an aside we mention that the *nonlinear* front solution of Nozaki and Bekki [Eqs. 2-7;39] yields for our GLE a front that is not stationary at $\mu_{conv}^c$ but rather moves with velocity

$$v_{NB} = -(3\sqrt{\frac{1+c_1^2}{8+9c_1^2}} - 1)v_g \qquad (37)$$

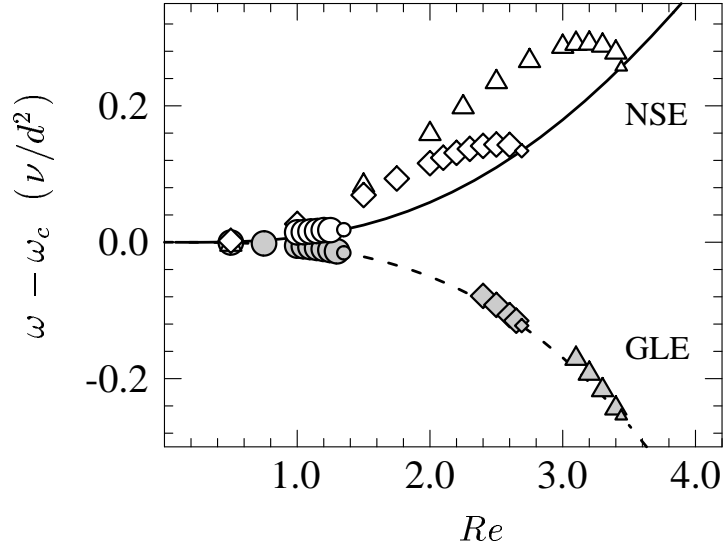

**Fig. 4.** Selected oscillation frequency of PV flow in the Taylor-Couette system obtained from the NSE (open symbols) and GLE (half-tone symbols) versus Reynolds number. The full (dashed) line is the front frequency $\Omega_F = \Re\zeta(Q_s)$ of the linearized NSE (GLE) at the border between absolute and convective instability reduced by the viscous diffusion rate $\nu/d^2$ across the gap. Parameters are $\epsilon = 0.0288$ (circles), $\epsilon = 0.114$ (diamonds), and $\epsilon = 0.186$ (triangles). The small symbols marking $\Omega_F$ for these three $\epsilon$ values show that the eigenfrequencies of the nonlinear equations in the absolutely unstable regime approach the $\Omega_F$ limit curve at the right places.

in the upstream direction with a bulk wave number

$$\xi_0(q_b)_{NB} = \frac{\sqrt{\mu^c_{conv}}}{(c_2-c_1)\sqrt{8+9c_1^2}} \times$$
$$\left[3(1+c_1^2) + sign(c_2-c_1) \times \sqrt{8(c_2-c_1)^2 + 9(1+c_1c_2)^2}\right]. \qquad (38)$$

So we conclude that the *nonlinear* front solution of Nozaki and Bekki is unrelated to the PV patterns at $\mu^c_{conv}$. Furthermore, it was noted [40] that localized initial perturbations did not evolve into the *nonlinear* front solution of Nozaki and Bekki [39, 41, 42]. However, $v_{NB}$ and $(q_b)_{NB}$ differ for small through-flow only slightly from the respective values $v_F = 0$ and $(q_b)_F$. For example, at $Re = 2$ one obtains $v_{NB} = -0.06 v_g$ and $(q_b)_{NB} = 1.15 (q_b)_F$.

**Scaling properties** Scaling length and time according to

$$\hat{z} = \sqrt{\mu}\frac{z}{\xi_0} \; ; \; \hat{t} = \mu\frac{t}{\tau_0}, \qquad (39)$$

the GLE for the reduced amplitude $\hat{A} = A\sqrt{\gamma/\mu}$ no longer contains $\gamma$ and $\mu$ explicitly but the reduced group velocity $V_g$ (7) and the coefficients $c_0, c_1, c_2$. Thus the reduced selected frequency $\hat{\Omega} = \tau_0 \Omega/\mu$ and the selected bulk wave number $\hat{q}_b = \xi_0 q_b/\sqrt{\mu}$ depend not only on $V_g$ but via $c_0, c_1, c_2$ also on the Reynolds number. This $Re$-dependence can alternatively be seen — via the $Re$-dependence of $V_g$ and of the $c_i$'s entering (7) — also as an additional $\mu$-dependence. The latter dependence is sufficiently strong to prevent a scaling of $\hat{\Omega}$ with $V_g$ alone. However, by reducing the eigenfrequencies $\Omega(\epsilon, Re)$ selected for our three different $\epsilon$-values by the limiting values $\Omega(\epsilon, Re^c_{conv}) = \Omega^c_{conv}$ (27) at the border of the absolutely unstable regime we effectively have eliminated the $\mu$-dependence and all GLE data (half-tone symbols in Fig. 5) fall onto one curve. A similar scaling holds for the bulk wave number $q_b$ divided by $(q_b)^c_{conv}$ (28).

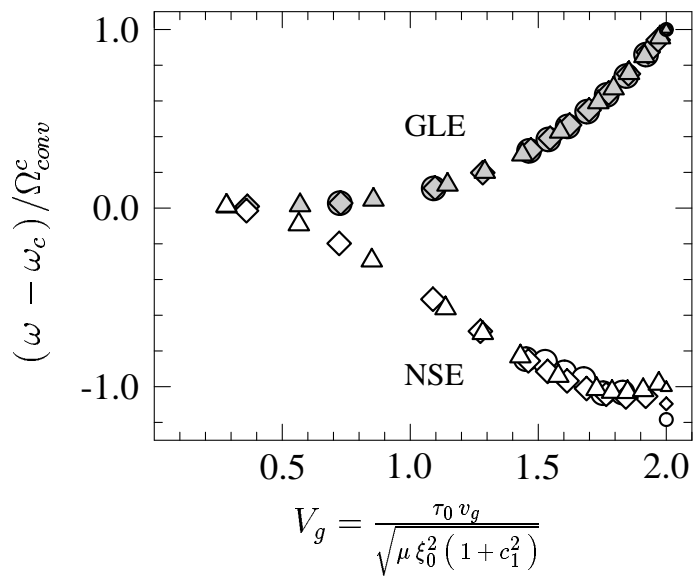

**Fig. 5.** Frequency shift $\omega - \omega_c$ of PV flow in the Taylor-Couette system versus scaled group velocity $V_g$ (7). Both, NSE (open symbols) and GLE (half-tone symbols) results are scaled by the corresponding GLE frequency $\Omega^c_{conv}$ (27) at the convective instability border for the respective parameters $\epsilon = 0.0288$ (circles), $\epsilon = 0.114$ (diamonds), and $\epsilon = 0.186$ (triangles). The small symbols have the same meaning as explained in Fig. 4.

### 4.2 Pattern selection within the NSE

Remarkably enough, by reducing the NSE frequencies with the GLE frequency $\Omega^c_{conv}$ (27) the NSE results (open symbols in Fig. 5) almost show this one-variable scaling with $V_g$. However, for the GLE $(\omega - \omega_c)/\Omega^c_{conv}$ increases monotonously with the scaled group velocity $V_g$ whereas the NSE results show a monotonous decrease with $V_g$. This discrepancy between NSE and GLE results is caused by the different dispersion relations of the full hydrodynamic equations and the approximate GLE — cf. below.

The NSE wavelengths that are selected when the basic state is enforced at the inlet and outlet weakly decrease with increasing $Re$ and $\epsilon$. For this BC the wavelengths are mostly smaller than the critical one with a deviation of up to 2% for the largest $\epsilon = 0.186$ that was investigated [18]. Qualitative and quatitative similar results were obtained for transversal Rayleigh-Bénard convection rolls in horizontal shear flow [15, 16]. These deviations are much stronger than the decrease of the critical wavelength with increasing Reynolds numbers. To summarize: the wavelengths resulting from the NSE differ systematically from those resulting from the GLE. The former decrease with growing $Re$ and $\epsilon$ while the latter increase. Also experiments [43, 44] yield wave numbers which differ in a to our NSE results common distinctive way from the proper GLE result (36). Nevertheless we think that the selection mechanism is quite similar — c.f. the discussion below.

### 4.3 Comparison with front propagation — NSE *versus* GLE

Let us first consider Taylor vortices without through-flow. Then the frequency $\Omega$ is zero, the GLE contains no imaginary coefficients $c_i$, and the first-order spatial derivative is absent since $v_g = 0$. In this case the GLE yields for propagating fronts [45] as well as for stationary patterns in finite systems [38] the critical wave number, $q(z) = 0$, all over the extension of the pattern. The BC $A = 0$ enforces the collaps of the supercritical band of stable bulk wave numbers of *nonlinear* vortex patterns to $q = 0$. This *nonlinearly* selected wave number $q = 0$ happens to be for $Re = 0$ the same as the wave number of maximal *linear* growth under a *linear* front whose spatiotemporal evolution is governed by the dispersion $\zeta_{GLE}(Q, \epsilon; Re = 0)$ (25) of the *linear* GLE. In the presence of through-flow, however, the wave number of maximal growth under the *linear* front — e.g. $q_F$ (34) — of the GLE differs from the one in the *nonlinear* bulk far behind the front $[(q_b)_F$ (36)].

Now, already without through-flow the dispersion relation of the *linear* NSE differs significantly from that of the GLE so that the wave number of largest temporal growth under a *linear* NSE front deviates from the GLE result $q = 0$. Moreover it has been observed experimentally [3, 10] and numerically for Taylor vortices [6, 46] or Rayleigh-Bénard convection rolls [11] that the wave number of real fronts is spatially varying in contradiction to the GLE picture.

Now let us consider the through-flow case with a nonvanishing frequency eigenvalue $\Omega$. Then also in the GLE approximation the selected flow patterns reveal spatially dependent wave number profiles connecting the wave number of maximal growth in the far tail of the front or at the inlet to the one in the bulk. In Fig. 4 we compare the eigenfrequencies of the *nonlinear* NSE (open symbols) with those (solid line) that would be selected by a *linear* front that is stationary at the border of the absolutely unstable regime. To that end Recktenwald and Dressler [35] have determined the dispersion relation $\zeta(Q)$ of linear perturbations $\sim e^{i[(k_c+Q)z-(\omega_c+\zeta)t]}$ in the complex $Q$-plane resulting from the NSE with a shooting method as described in Ref. [28]. The solid line in Fig. 4 represents the frequency $\Omega_F = \Re\zeta(Q_s)$ selected by the stationary front ($v_F = 0$) of the *linear* NSE. Again, as discussed for the GLE in Sec. 4.1, this frequency is determined by the saddle $Q_s$ of $\zeta(Q)$ for which the temporal growth rate $\Im\zeta(Q_s) = 0$. This defines the boundary between absolute and convective instability. With increasing $Re$ the eigenfrequency of the *nonlinear* NSE approaches the frequency selected by the *linear* stationary front at the border of the absolutely unstable regime as indicated for the three $\epsilon$-values shown in Fig. 4. The front frequencies for these $\epsilon$-values are marked by small open symbols on the curve $\Re\zeta(Q_s)$ to indicate that the eigenfrequencies of the NSE in the absolutely unstable regime (large open symbols in Fig. 4) do indeed end at the right positions on this curve.

This NSE behavior is very similar to the GLE behavior (half-tone symbols and dashed line). Note, however, that the linear dispersion relations of NSE and GLE are different: the front frequency $\Re\zeta_{NSE}(Q_s)$ (solid line in Fig. 4) is positive and increases with through-flow while $\Re\zeta_{GLE}(Q_s)$ (dashed line) is negative and decreases with through-flow. This difference in the dispersion relations seems to be the major cause for the differences in the selected patterns.

## 5 Conclusion

We have reviewed propagating vortex structures in systems of finite length that propagate downstream in an externally applied flow. Within the absolutely unstable parameter regime a unique pattern selection is observed. The selected PV patterns are independent of parameter history, initial conditions, and system length provided the latter is large enough to allow for a saturated bulk region with homogeneous pattern amplitude and wave number. But they depend on the inlet and outlet BC. The outlet BC influences the structures only locally in the vicinity of the outlet while the globally constant frequency and the streamwise profiles of pattern amplitude and wave number and their bulk values are basically determined by the inlet BC. The selected frequency $\omega$ of the pattern oscillation is an eigenvalue of a nonlinear eigenvalue problem with the set of time periodic fields being the associated eigenfunction. The analysis of the appropriate GLE approximation shows that the eigenfunction associated to the selected eigenfrequency varies as smoothly and as little as possible under the imposed BC. This property is similar to the ground state behavior of the linear stationary quantum mechanical Schrödinger equation.

Different lateral BCs entail different frequencies and eigenfunctions, i.e., lateral pattern profiles. For example, inlet conditions that exert a phase pinning force on one or more of the fields of traveling vortices reduce the frequency. The spatial growth behavior and the pattern profile of the PV structures in the region between inlet and bulk saturated pattern is strongly influenced by the inlet BC. But with increasing growth length of the pattern from the inlet the influence of the latter on spatiotemporal properties of the convection pattern becomes weaker. While the frequencies and bulk wave numbers obtained for various BCs and through-flow rates well inside the absolutely unstable regime differ substantially they all fall onto a linear dispersion curve. Finally, when approaching with increasing $Re$ the border between absolute and convective instability the eigenfrequencies and also the spatial pattern profiles belonging to different BCs approach each other. On this border the eigenvalue problem becomes effectively linear and the pattern selection mechanism becomes that one of linear front propagation.

### Acknowledgments

Support by the Deutsche Forschungsgemeinschaft and the Stiftung Volkswagenwerk is gratefully acknowledged.